\documentclass{article}
\usepackage{ijcai26}    

\usepackage{times}
\usepackage{soul}
\usepackage{url}
\usepackage[hidelinks]{hyperref}
\usepackage[utf8]{inputenc}
\usepackage[small]{caption}
\usepackage{graphicx}
\usepackage{amsmath}
\usepackage{amsthm}
\usepackage{booktabs}
\usepackage{multirow}
\usepackage{tabularx}
\usepackage{algorithm}
\usepackage{algorithmic}
\usepackage[switch]{lineno}
\usepackage{tcolorbox}
\tcbuselibrary{skins,breakable}
\usepackage{listings}
\usepackage{array}
\usepackage{subcaption}
\usepackage{amssymb}
\usepackage[table]{xcolor}

\definecolor{seccol}{RGB}{225,240,225}

\title{HardSecBench: Benchmarking the Security Awareness of LLMs for Hardware Code Generation}

\author{
Qirui Chen$^1$ \and Jingxian Shuai$^1$ \and Shuangwu Chen$^{1}$\thanks{\ Corresponding author.} \and Shenghao Ye$^1$ \and Zijian Wen$^1$ \and Xufei Su$^2$ \and \\
Jie Jin$^3$ \and Jiangming Li$^4$ \and Jun Chen$^4$ \and Xiaobin Tan$^1$ \And Jian Yang$^1$
\\
\affiliations
$^1$University of Science and Technology of China, China \\
$^2$Xi'an Jiaotong University, China \\
$^3$Nanjing University, China \\
$^4$ZTE Corporation, China
\\
\emails
\{chenqirui, vexertron\_shuai\}@mail.ustc.edu.cn,
chensw@ustc.edu.cn
}

\begin{document}

\maketitle


\begin{abstract}
Large language models (LLMs) are increasingly used for hardware and firmware code generation, but existing studies primarily evaluate functional correctness while largely overlooking security. 
However, LLM-generated code that appears functionally sound may embed security flaws which could induce catastrophic damages after deployment. 
This critical research gap motivates us to design a benchmark for assessing security awareness under realistic specifications. 
In this work, we introduce \textbf{HardSecBench}, a benchmark with 924 tasks spanning Verilog Register Transfer Level~(RTL) and firmware-level C, covering 76 hardware-relevant Common Weakness Enumeration~(CWE) entries. Each task includes a structured specification, a secure reference implementation, and executable tests. To automate artifact synthesis, we propose a multi-agent pipeline that decouples synthesis from verification and grounds evaluation in execution evidence, enabling reliable evaluation.
We evaluate diverse LLMs and find that they often satisfy functional requirements while leaving security risks. 
We also find that security results vary with prompting. These findings highlight pressing challenges and offer actionable insights for future advancements in LLM-assisted hardware design. Our data and code are available at \url{https://github.com/chenqirui2002/HardSecBench}.
\end{abstract}

\section{Introduction}

\begin{figure}[t]
    \centering
    \includegraphics[width=0.95\linewidth]{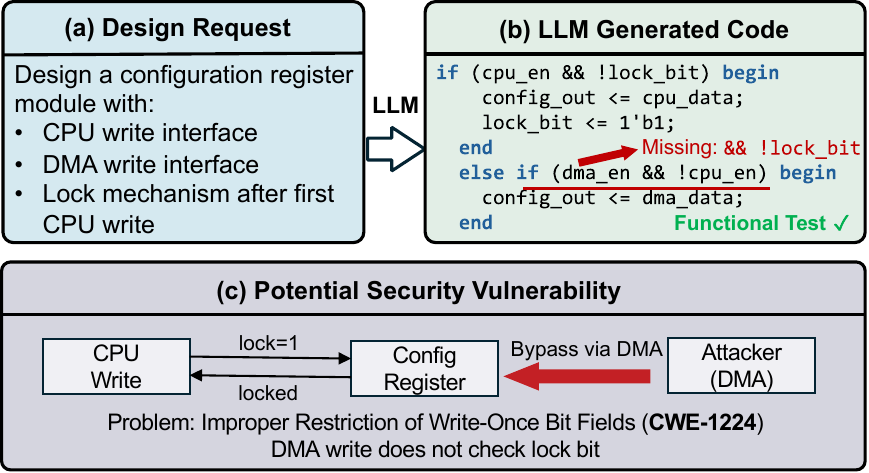}
    \caption{Potential security vulnerability in LLM-generated code. The LLM produces functionally correct code~(b) from a design request~(a), but the DMA interface fails to check the lock status, enabling attackers to bypass write-once protection~(c).}
    \label{fig:introcase}
\end{figure}

Large Language Models~(LLMs) have been increasingly incorporated into practical hardware and firmware development for automated code generation, including Register Transfer Level (RTL) design and C-based firmware development~\cite{liu2024chipnemodomainadaptedllmschip,11105957}. 
Extensive studies have focused on improving and evaluating the functional correctness of LLM-generated code~\cite{10323812,lu2024rtllm}.
However, as security intent is often not explicitly specified in LLM-assisted hardware development,  the generated codes may pass functional validation yet still harbor potential security vulnerabilities that may induce catastrophic damages~\cite{10.1145/3576915.3623157,ijcai2025p895}. For example, in Figure~\ref{fig:introcase}, an LLM-generated configuration register   successfully passes read/write tests, but   violates  the write-once lock policy when an alternative access interface fails to enforce the lock. As designs scale up and interfaces proliferate, such security  violations may accumulate over the entire development lifecycle, thereby exacerbating systemic risks.


However, there is still a lack of standardized and comprehensive benchmark to evaluate the systematic security of LLM-based hardware designs, as existing studies ~\cite{ijcai2025p895,hu2025secfsmknowledgegraphguidedverilog,long2025veriloglavdllmaidedrulegeneration} usually validate the security of LLM-generated code on their own specific test sets. 
Two critical challenges hinder the development of a unified benchmark for systematic security assessment.


\paragraph{Trustworthy  benchmark construction at scale.}
High-quality security benchmark samples are expensive to build and validate, making it difficult to expand evaluation to large benchmarks through expert effort alone~\cite{10.1145/3670474.3685956}. 
To mitigate this inefficiency, LLM-assisted synthesis offers a practical way to generate benchmark tasks in large numbers. 
However, ensuring the reliability of the resulting benchmark and evaluation remains challenging, requiring rigorous quality control to keep results trustworthy~\cite{zhu2025qimengcodevr1}. 

\paragraph{Unreliable security evaluation.}
Existing RTL security analysis workflows often rely on experts to craft SystemVerilog Assertions~(SVAs) from security intent and debug violations through simulation~\cite{DBLP:conf/dac/Orenes-VeraMWM21,afsharmazayejani2024toward}. However, SVAs can miss vulnerabilities when security-relevant behaviors are not triggered.
To reduce reliance on simulation scenarios, some workflows use early-stage static rule checking, which still demands security expertise and often emphasizes functional issues over security properties~\cite{synopsys_spyglass}.
Recent work also adopts LLM-as-a-judge verifiers, but without timing-accurate simulation evidence their judgments can be unreliable and inconsistent across runs~\cite{bhatt2024cyberseceval2widerangingcybersecurity}.

To address these challenges, we design an automated pipeline that scales benchmark construction and enables security evaluation under specifications that do not reveal security intent.
Using this pipeline, we build HardSecBench: each task includes a structured specification that separates functional and security requirements, and requirement-level harnesses that validate requirements via simulation or execution, with tests that trigger security-relevant behaviors.
This pipeline reconciles mismatches until the specification, implementation, and harnesses agree on all checks, and we then apply coverage and mutation filtering to retain harnesses that better distinguish correct from incorrect implementations.
On the evaluation side, HardSecBench avoids subjective judging and scores security using simulation evidence from targeted harnesses that actively trigger security relevant behaviors.
The benchmark covers 76 Common Weakness Enumeration entries~(CWE)~\cite{mitre-cwe} and evaluates whether models implement protections checked by the security requirements.
Finally, HardSecBench reports results under single-attempt generation and iterative refinement, where a collaborator helps repair functional defects so security tests run on a correct functional baseline and reduce confounding.

Based on HardSecBench, a comprehensive evaluation of state-of-the-art~(SOTA) models reveals that many models can generate functionally valid hardware designs yet lack critical security protections.
To explore the security potential of LLMs, we conduct a prompt sensitivity analysis and find that these SOTA models could yield notable security gains even with merely generic security reminders.
This finding indicates that the security knowledge is inherently embedded in LLMs but has not been fully exploited.
Accordingly, providing explicit security guidance can effectively compensate for the flaws in LLM-generated code.
We further conduct an empirical study on specialized models to evaluate the impact of domain-specific training on their security performance.
Finally, we carry out a systematic analysis of vulnerability to highlight common weaknesses across diverse models.
 
 

The primary contributions of this work are as follows:
\begin{itemize}
    \item We formulate a systematic evaluation setting for security awareness in LLM-generated hardware designs, offering a rigorous testbed for secure hardware code generation;
    \item We develop a multi-agent construction pipeline that synthesizes benchmark samples, producing testable specifications and corresponding simulatable harnesses;
    \item We propose HardSecBench, a benchmark of 924 Verilog and firmware-C tasks spanning 76 CWE entries;
    \item Using HardSecBench, we comprehensively evaluate a wide range of SOTA models and report our findings on the security risks of LLM-generated hardware designs.
\end{itemize}

\section{Related Work}
\subsection{LLM-Assisted Hardware Design}
LLM-assisted hardware design has been studied across the RTL development lifecycle, including Verilog generation from specifications and interactive co-design that supports iterative refinement and debugging~\cite{10137086,10299874,zhu2025qimengcodevr1}. Simulation-driven benchmarks have been adopted to standardize evaluation by compiling and simulating generated RTL against reference testbenches, as exemplified by VerilogEval and RTLLM~\cite{10323812,10.1145/3718088,lu2024rtllm}. Recent efforts release larger datasets and lightweight fine-tuned LLMs to reduce the training and inference costs for adopting RTL generation, enabling more accessible deployment in local environments~\cite{10720939,zhu2025qimengcodevr1}. Beyond module synthesis, domain-adapted assistants and agentic EDA systems extend LLM support to engineering question answering, script generation and partial flow automation~\cite{liu2024chipnemodomainadaptedllmschip,wu2024chateda}. LLMs have also been explored for generating verification items, including test inputs and assertions, along with benchmarks to assess these outputs~\cite{10691792,pulavarthi-etal-2025-assertionbench}.

\subsection{Security of LLM-Generated Code}
Security evaluation of LLM-generated artifacts has been studied extensively in software, showing that code can satisfy functional specifications while still containing potential vulnerabilities that are not exercised by standard requirements~\cite{peng2025cwevaloutcomedrivenevaluationfunctionality,li2025safegenbenchbenchmarkframeworksecurity,kim2025rethinking,10.1145/3711896.3736561}. 
In addition to evaluation, software engineering research proposes methods that steer generation toward secure implementations. Constrained decoding enforces security constraints during generation to reduce insecure patterns~\cite{fu2024constrained}. Retrieval augmented generation adds selected security knowledge and secure coding patterns to guide implementation choices~\cite{shi2025rescueretrievalaugmentedsecure}. 
In hardware settings, recent work examines risks including intellectual property leakage and memorization~\cite{mashnoor2025circuitguardmitigatingllmmemorization}, as well as hardware hallucinations in generated designs~\cite{11106012}. Knowledge-augmented approaches incorporate structured domain knowledge to guide security-aware synthesis or enable automated structural auditing through graph-based traversal rules~\cite{ijcai2025p895,hu2025secfsmknowledgegraphguidedverilog,long2025veriloglavdllmaidedrulegeneration}.

\section{HardSecBench}
\label{sec:construction}

\begin{figure*}[t]
  \centering
  \includegraphics[width=0.99\textwidth]{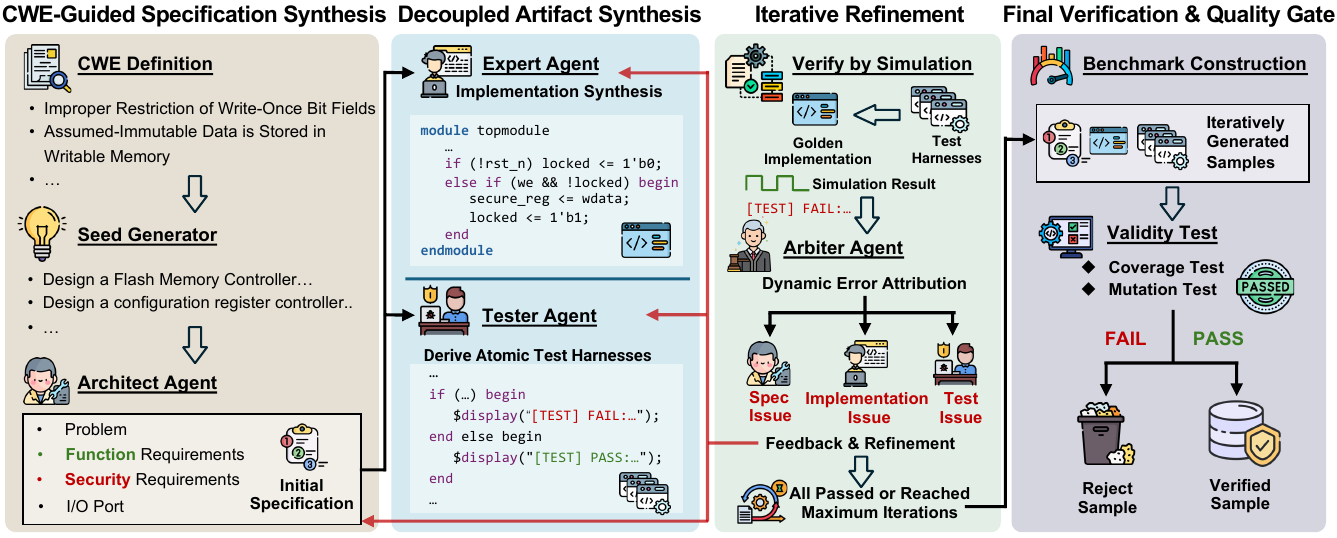}
  \caption{HardSecBench construction pipeline. From CWE-derived seeds, the Architect builds a structured specification $P_i$ that separates functional requirements $\mathcal{R}^{f}_{i}$ from security requirements $\mathcal{R}^{s}_{i}$, so implementations can be elicited from $\mathcal{R}^{f}_{i}$ while security is checked against $\mathcal{R}^{s}_{i}$. Given $P_i$, the Expert and Tester independently synthesize the golden implementation and requirement-level harnesses, which are run and iteratively reconciled until all checks agree; a final quality gate filters retained benchmark samples.}
  \label{fig:pipeline}
  \vspace{-5pt}
\end{figure*}

In this section, we introduce HardSecBench, a benchmark to evaluate hardware security awareness via a multi-agent construction pipeline.
As shown in Figure~\ref{fig:pipeline}, the workflow has four stages coupled by a single structured specification $P_i$. 
Starting from CWE-derived seeds, the Architect produces $P_i$ and separates functional requirements $\mathcal{R}^{f}_{i}$ from security requirements $\mathcal{R}^{s}_{i}$, so that the functional specification can be used to elicit implementations without revealing security intent while $\mathcal{R}^{s}_{i}$ defines what security protections will be checked.
Given the same $P_i$, the Expert and Tester synthesize the golden implementation and requirement-level harnesses in separate branches to avoid logic coupling between implementation and verification.
We then execute the harnesses against the implementation and iteratively reconcile mismatches until all requirement checks agree. We finally apply a quality gate, retaining only samples whose harnesses provide strong diagnostic signals.

\subsection{CWE-Guided Specification Synthesis}
We use a hierarchical synthesis strategy to promote diversity and maintain specification quality.
The process starts with seed synthesis, our Seed Generator converts each CWE definition into a seed that specifies the implementation language and a 1--2 sentence scenario that the target weakness can arise.
The Architect agent then expands each seed into a structured specification $P_i$ for task $i$, including a problem statement, an I/O interface specification, and two requirement sets: functional requirements $\mathcal{R}^{f}_{i}$ and security requirements $\mathcal{R}^{s}_{i}$.
By design, $\mathcal{R}^{f}_{i}$ states only functional behavior and avoids revealing security intent, while $\mathcal{R}^{s}_{i}$ lists the protections required by the CWE guidance.
This separation serves two purposes: during construction, we keep $\mathcal{R}^{f}_{i}$ free of security content, and during evaluation, we give the target model without $\mathcal{R}^{s}_{i}$, while we use $\mathcal{R}^{s}_{i}$ to evaluate its security awareness.

\subsection{Decoupled Artifact Synthesis}
To ensure objectivity and prevent logic coupling, we enforce strict information isolation between golden-implementation synthesis and test-harness generation.
Without isolation, harnesses may inadvertently encode implementation details, inflating pass rates and weakening diagnostic validity.
The Expert and Tester agent follow independent generation paths and use $P_i$ as their only shared reference.
The Expert synthesizes the golden implementation to satisfy both $\mathcal{R}^{f}_{i}$ and $\mathcal{R}^{s}_{i}$, while the Tester derives atomic harnesses with a one-to-one mapping to requirements in $\mathcal{R}^{f}_{i}$ and $\mathcal{R}^{s}_{i}$.
Each harness is self-contained and takes the form of a testbench for RTL tasks or a standalone C driver for firmware tasks.
The test harness emits a standardized \texttt{PASS/FAIL} trace so that verification can parse results deterministically.
Both branches iteratively refine their initial outputs until they reach a compilable state. This setup ensures the code and harnesses can compile and run before refinement, so later failures reflect requirement mismatches rather than basic build or runtime errors.

\subsection{Arbiter-Driven Iterative Refinement}
We aim to improve sample yield by repairing correctable mismatches instead of discarding a task as soon as a requirement check fails.
A failing check can be caused by an ambiguous requirement, an implementation error, or a harness bug, since the three artifacts are synthesized in separate branches.
The Expert, the Tester, and the Architect each operate under restricted views of the artifacts, so none of them can reliably determine whether the failure is due to the specification, the implementation, or the harness.
This motivates an Arbiter-coordinated loop.
The Arbiter observes the structured specification $P_i$, the golden implementation, the requirement-level harnesses, and runtime evidence from execution and analysis traces produced by the harnesses, and uses them to localize the source of the mismatch.
It then issues targeted feedback that specifies what to revise and which observed evidence supports the revision.
We repeat diagnosis, revision, and re-execution until all requirement checks pass, establishing consistency among the specification, the golden implementation, and the harnesses.

\subsection{Final Verification and Quality Gate}
Prior to benchmark inclusion, we apply a quality gate to reduce false negatives caused by weak harnesses.
Without this gate, a harness may run but still miss security relevant behaviors, allowing insecure implementations to pass.
We screen samples using two complementary signals.
First, a coverage signal rules out cases where the harness exercises only a small portion of the design.
Second, a mutation signal checks whether the harness can distinguish the intended implementation from perturbed variants that introduce deviations.
In addition to these automated checks, domain experts manually audited 100 randomly sampled tasks and found that the artifacts largely satisfy HardSecBench's quality criteria.
Together, they improve the reliability and diagnostic power of the benchmark for evaluating hardware security awareness.

\section{Evaluation Methodology} 

\subsection{Evaluation Settings}
\begin{figure}[t]
  \centering
  \includegraphics[width=0.43\textwidth]{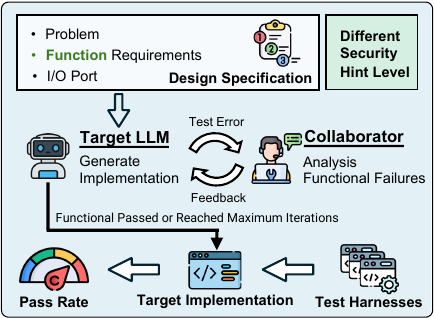}
 \caption{HardSecBench evaluation architecture. The target model generates an implementation, the Collaborator provides feedback for fixing functional issues.}
  \label{fig:eval}
  \vspace{-5pt}
\end{figure}

Our evaluation focuses on how the target model generates an implementation under different security hint settings, with an optional Collaborator that provides feedback on functional issues as shown in Figure~\ref{fig:eval}.
We evaluate all models using the requirement-level harnesses associated with each HardSecBench sample.
Each harness emits a \texttt{PASS/FAIL} marker per requirement, which we parse to compute requirement-level pass rates. 
We then define two evaluation settings: one that evaluates the model after a single generation attempt, and one that allows iterative revisions with functional feedback before the final security evaluation to separate security awareness from functional defects.

\paragraph{Single-Attempt Evaluation.}
This setting measures the model's security intuition by checking whether it adds protections beyond functional correctness when it receives only functional requirements and no execution feedback.
We require a compilable implementation so that syntax errors do not dominate the security measurement.
If the initial output fails to compile, we allow up to three rounds of compilation and fixing using only compiler error messages, and we evaluate the first version that successfully compiles.
These fixes aim only to make the code runnable and do not reveal outcomes from functional or security harnesses.
After a compilable implementation is obtained, we run the full functional and security harnesses once and record the results without providing them back to the model.

\paragraph{Iterative Refinement Evaluation.}
This setting simulates a collaborative workflow that improves functional correctness before assessing security.
In each iteration, the model produces a compilable implementation, we run the functional harnesses, and the Collaborator provides feedback on functional failures and code issues.
The Collaborator provides functional-only feedback and does not comment on security.
We repeat iterations until all functional requirements pass or the iteration limit is reached, and then evaluate security performance on the final implementation.
We evaluate security only after the implementation reaches a fully functional or near-complete state, which yields a cleaner measure of security awareness without confounding from functional defects.

\subsection{The Pass@k Metric}
To quantify generation stability across independent samples, we report Pass@k for both functional and security requirements using a unified definition.
For each task $P_i$, we start from the requirement sets $\mathcal{R}^{f}_{i}$ and $\mathcal{R}^{s}_{i}$ and further decompose them into atomic requirements, each of which is checked by a requirement-level harness and serves as one evaluation unit.
Let $d \in \{\mathrm{Func}, \mathrm{Sec}\}$ denote the requirement type, where $\mathrm{Func}$ corresponds to $\mathcal{R}^{f}_{i}$ and $\mathrm{Sec}$ corresponds to $\mathcal{R}^{s}_{i}$.
Let $N$ be the number of tasks.
For task $P_i$ and type $d$, let $M_{i,d}$ be the number of atomic requirements derived from $\mathcal{R}^{d}_{i}$.
We draw $n$ independent generations for each task, where $n$ is the number of sampled candidate solutions per task. For each atomic requirement $j \in \{1,\dots,M_{i,d}\}$, let $c_{i,d,j}$ be the number of generations~(out of $n$) that pass this requirement.
We compute Pass@k under sampling without replacement for each atomic requirement and then average over all atomic requirements of type $d$ across tasks:
\begin{equation}
R_{d}@k
= \frac{1}{\sum_{i=1}^{N} M_{i,d}}
\sum_{i=1}^{N}
\sum_{j=1}^{M_{i,d}}
\left( 1 - \frac{\binom{n - c_{i,d,j}}{k}}{\binom{n}{k}} \right).
\end{equation}
This aggregation gives equal weight to each atomic requirement, which matches our requirement-level harness design.
We report $R_d@k$ for $k \le n$.

\newcolumntype{L}[1]{>{\raggedright\arraybackslash\hsize=#1\hsize}X}
\newcolumntype{C}[1]{>{\centering\arraybackslash\hsize=#1\hsize}X}

\begin{table*}[th]
    \centering
    \small
    \setlength{\tabcolsep}{4pt}
    \renewcommand{\arraystretch}{1.05}

    \begin{tabularx}{\textwidth}{
        L{1}
        C{0.4}
        C{0.33}
        C{0.33}
        C{0.33}
        C{0.33}
        C{0.33}
        C{0.33}
        C{0.33}
        C{0.33}
    }
        \toprule
        \multirow{2}{*}{\textbf{Models}} &
        \multirow{2}{*}{\textbf{SWE-bench}} &
        \multicolumn{2}{c}{\textbf{VerilogEvalV2}} &
        \multicolumn{2}{c}{\textbf{HSBench$_{\text{1-attempt}}$@1}} &
        \multicolumn{2}{c}{\textbf{HSBench@1}} &
        \multicolumn{2}{c}{\textbf{HSBench@5}} \\
        \cmidrule(lr){3-4} \cmidrule(lr){5-6} \cmidrule(lr){7-8} \cmidrule(lr){9-10}
        & & Pass@1 & Pass@20
          & Func. & \cellcolor{seccol}{Sec.}
          & Func. & \cellcolor{seccol}{Sec.}
          & Func. & \cellcolor{seccol}{Sec.} \\
        \midrule

        \multicolumn{10}{c}{\textit{Closed-source Models}} \\
        \midrule
        Claude-4.5-Opus      & 74.40 & 85.26 & 91.03 & 91.72 & \cellcolor{seccol}{37.45} & 97.11 & \cellcolor{seccol}{42.91} & 98.01 & \cellcolor{seccol}{45.66} \\
        Claude-4.5-Sonnet     & 70.60 & 75.00 & 85.26 & 87.86 & \cellcolor{seccol}{32.31} & 95.39 & \cellcolor{seccol}{38.68} & 98.88 & \cellcolor{seccol}{45.07} \\
        GPT-5.1-medium        & 66.00 & 83.33 & 92.31 & 91.68 & \cellcolor{seccol}\underline{40.02} & 97.27 & \cellcolor{seccol}{\underline{45.76}} & 98.76 & \cellcolor{seccol}{\textbf{53.88}} \\
        GPT-5-medium          & 65.00 & 83.33 & 91.03 & 91.35 & \cellcolor{seccol}{39.03} & 95.98 & \cellcolor{seccol}{42.96} & 98.82 & \cellcolor{seccol}{50.42} \\
        o1                    & -     & 73.72 & 85.90 & 89.15 & \cellcolor{seccol}{35.57} & 96.31 & \cellcolor{seccol}{41.66} & 98.97 & \cellcolor{seccol}{49.66} \\
        Gemini-3-Pro-Preview  & 74.20 & 85.90 & 93.59 & 93.93 & \cellcolor{seccol}{\textbf{42.51}} & 97.71 & \cellcolor{seccol}{\textbf{46.11}} & 99.08 & \cellcolor{seccol}{\underline{52.49}} \\
        Gemini-2.5-Pro        & 53.60 & 81.41 & 91.67 & 91.43 & \cellcolor{seccol}{39.29} & 96.57 & \cellcolor{seccol}{44.56} & 98.72 & \cellcolor{seccol}{51.51} \\
        Gemini-2.5-Flash      & 28.73 & 71.79 & 90.38 & 89.85 & \cellcolor{seccol}{32.07} & 95.68 & \cellcolor{seccol}{37.07} & 98.75 & \cellcolor{seccol}{44.46} \\
        Qwen3-Max             & 69.60 & 64.10 & 78.21 & 84.13 & \cellcolor{seccol}{30.70} & 93.17 & \cellcolor{seccol}{37.13} & 98.11 & \cellcolor{seccol}{44.58} \\
        
        \midrule
        
        \multicolumn{10}{c}{\textit{Open-source Models}} \\
        \midrule
        DeepSeek-V3.2         & 60.00 & 69.23 & 89.10 & 85.33 & \cellcolor{seccol}{32.11} & 97.03 & \cellcolor{seccol}{42.42} & 99.26 & \cellcolor{seccol}{49.00} \\
        Qwen3-Coder-480B-A35B & 55.40 & 63.46 & 76.28 & 85.20 & \cellcolor{seccol}{32.74} & 95.71 & \cellcolor{seccol}{39.74} & 98.58 & \cellcolor{seccol}{45.81} \\
        GLM-4.6               & 55.40 & 66.03 & 82.05 & 86.18 & \cellcolor{seccol}{\underline{35.67}} & 95.93 & \cellcolor{seccol}{\underline{42.89}} & 98.80 & \cellcolor{seccol}{\textbf{51.10}} \\
        Kimi-K2-Instruct-0905 & -     & 73.72 & 85.90 & 88.53 & \cellcolor{seccol}{33.06} & 96.18 & \cellcolor{seccol}{39.09} & 98.59 & \cellcolor{seccol}{45.03} \\
        GPT-OSS-120B          & 26.00 & 75.64 & 90.38 & 88.64 & \cellcolor{seccol}{34.07} & 96.33 & \cellcolor{seccol}{40.73} & 98.43 & \cellcolor{seccol}{46.21} \\
        Qwen3-Coder-30B-A3B   & -     & 48.08 & 62.18 & 73.70 & \cellcolor{seccol}{28.13} & 95.87 & \cellcolor{seccol}{39.63} & 98.39 & \cellcolor{seccol}{46.27} \\
        Llama-4-Maverick-Instruct & 21.04 & 55.13 & 66.67 & 84.72 & \cellcolor{seccol}{\textbf{35.87}} & 95.40 & \cellcolor{seccol}{\textbf{44.38}} & 98.05 & \cellcolor{seccol}{50.28} \\
        Llama-4-Scout-Instruct    & 9.06  & 37.18 & 50.64 & 77.12 & \cellcolor{seccol}{30.82} & 95.67 & \cellcolor{seccol}{40.66} & 98.20 & \cellcolor{seccol}{46.84} \\
        Qwen3-14B             & -     & 54.49 & 77.56 & 76.14 & \cellcolor{seccol}{30.05} & 95.31 & \cellcolor{seccol}{40.57} & 98.68 & \cellcolor{seccol}\underline{50.42} \\
        \bottomrule

    \end{tabularx}

    \caption{Main evaluation of LLMs under security settings. Func. and Sec. denote functional and security pass rates~(in \%). HSBench denotes HardSecBench; HSBench$_{\text{1-attempt}}$@1 denotes the single-attempt evaluation setting. Within each model group~(open-source and closed-source), \textbf{bold} indicates the best result and \underline{underline} indicates the second-best result.}
    \vspace{-5pt}
    \label{tab:main-eval}
\end{table*}

\section{Experimental Setup}
We evaluate a comprehensive suite of models, spanning closed-source models including Claude-4.5 Series~\cite{claude}, GPT-5 Series~\cite{gpt-5-system-card}, o1~\cite{gpt-o1}, Gemini Series~\cite{gemini} and Qwen3-Max~\cite{qwen3_model_card_2025}, alongside open-source models including DeepSeek-V3.2~\cite{deepseekai2024deepseekv3technicalreport}, GLM-4.6~\cite{glm_model_card_2025}, Kimi-K2-Instruct~\cite{kimi_model_card_2025}, Qwen3 Series~\cite{qwen_coder_model_card_2025}, Llama 4 Series~\cite{llama4} and GPT-OSS~\cite{openai2025gptoss120bgptoss20bmodel}. 
We report two metrics, Functional Pass Rate and Security Pass Rate, under an execution environment that uses \texttt{Icarus Verilog v12.0} for RTL designs and \texttt{gcc} for C implementations.
In the multi-iteration setting, the refinement loop is capped at 5 iterations. For each generated candidate, we allow up to 3 rounds of compile and fix.
API-based models are accessed via OpenRouter~\cite{openrouter} and SiliconFlow~\cite{siliconflow}, while local models are evaluated on NVIDIA A100-40G GPUs. We use GPT-5.1 to power the auxiliary agents that provide functional feedback. For the Collaborator agent, we set the temperature to 0.3 to keep its feedback stable. During benchmark construction, all agents are powered by Gemini-3-Pro-Preview.

\vspace{-2pt}
\section{Results and Analysis}

\subsection{Benchmark Statistics and Quality Validation}
\begin{figure}[t]
  \centering
  \begin{subfigure}[t]{0.49\columnwidth}
    \centering
    \includegraphics[width=\linewidth]{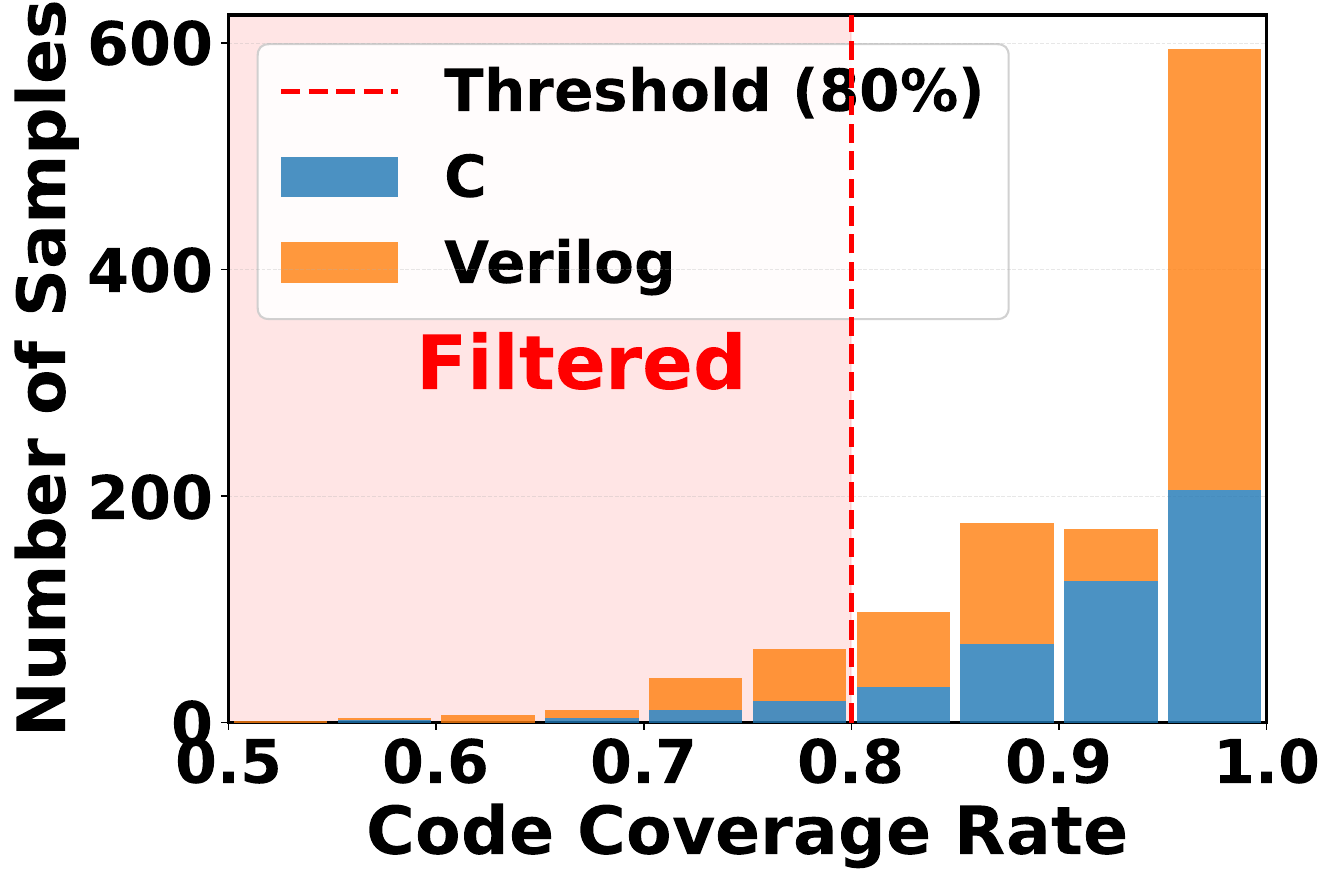}
    \caption{Coverage test distribution.}
    \label{fig:sub-test-a}
  \end{subfigure}\hfill
  \begin{subfigure}[t]{0.49\columnwidth}
    \centering
    \includegraphics[width=\linewidth]{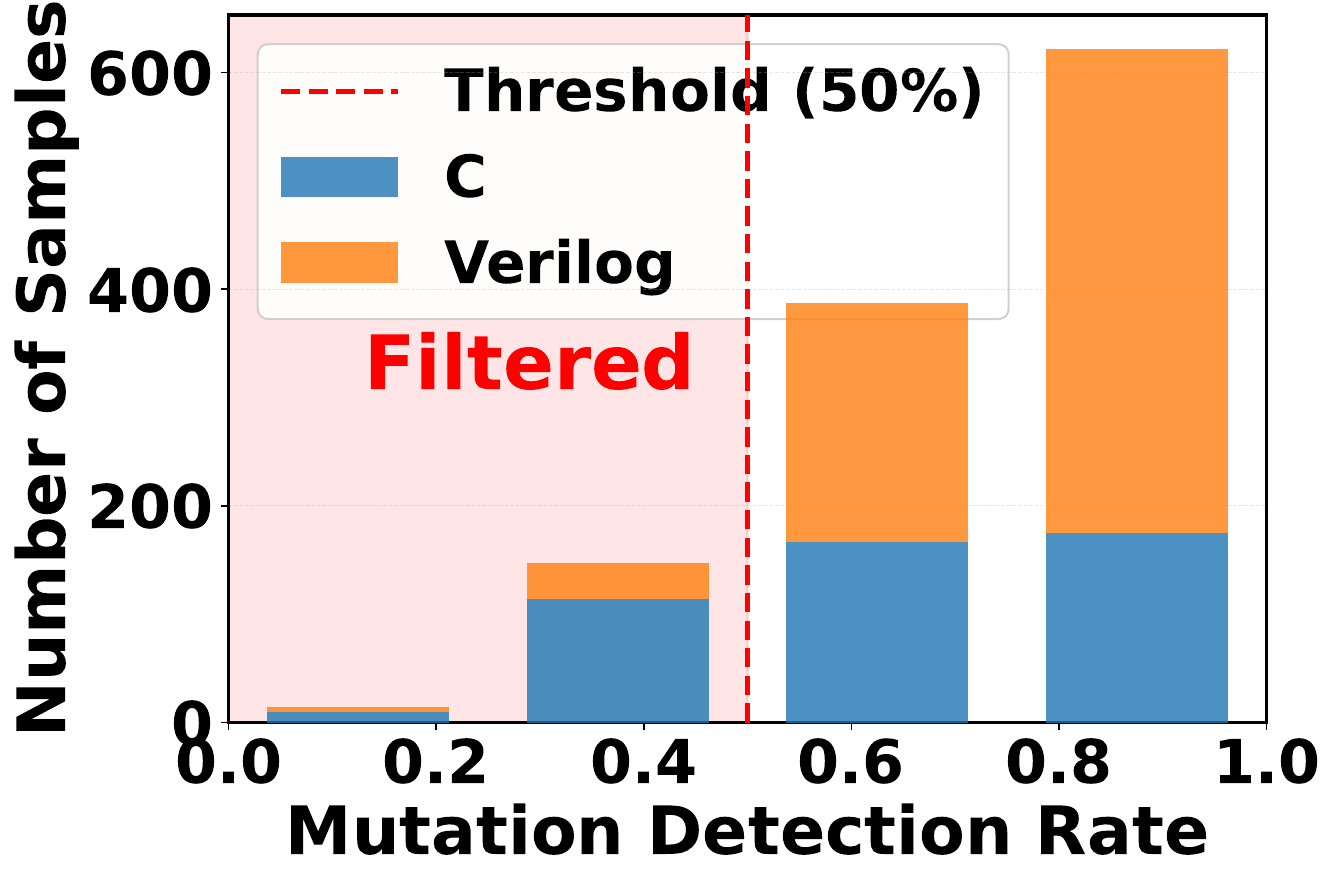}
    \caption{Mutation Test Distribution.}
    \label{fig:sub-test-b}
  \end{subfigure}
  \caption{Quality validation of HardSecBench:~(a) Code coverage and~(b) mutation detection distributions. Red dashed lines indicate the strict quality gates used for sample filtering.}
   \vspace{-5pt}
  \label{fig:test}
\end{figure}

Before evaluating the target models, we validate the correctness and internal consistency of HardSecBench.
We start from 1170 samples spanning 76 CWE entries and run our efficacy screening.
We remove 246 samples that do not meet our thresholds, yielding 924 validated tasks with 4425 test cases.
Among the retained tasks, 599 are Verilog~(64.8\%) and 325 are C~(35.2\%).
As shown in Figure~\ref{fig:test}, the retained tasks achieve high quality on two key dimensions.

\paragraph{Code Coverage Analysis.} To ensure the verification artifacts exercise meaningful paths, we impose a minimum coverage threshold of 80\%. For C artifacts, we measure line coverage using \texttt{gcov} with instrumentation. For Verilog, considering the varying support for automated coverage metrics in open-source toolchains, we estimate coverage via static analysis by mapping test harness signals to executable lines. As shown in Figure~\ref{fig:sub-test-a}, the benchmark attains 92.5\% average coverage, indicating strong internal consistency.

\paragraph{Discriminative Power via Mutation Analysis.} A test harness is useful only if it separates correct implementations from insecure variants. We apply five mutation operators: constant change, operator swap, condition negation, stuck-at signal, and assignment removal. For each task, we generate five mutants and mark detection if any test fails. We retain tasks only when the mutation score is at least 50\%. As shown in Figure~\ref{fig:sub-test-b}, the benchmark attains a mean mutation score of 70.2\%, indicating strong discriminative capability in exposing insecure implementations.

\subsection{Main Evaluation: Security Awareness}

We use reported SWE-bench Verified~\cite{jimenez2024swebench} results as a baseline for general software engineering capability and run VerilogEvalV2~\cite{10.1145/3718088} to measure RTL code generation capability.
On HardSecBench, we report single-attempt scores to reflect initial security awareness, and we report iterative-refinement scores together with Pass@5 to measure robustness under repeated generations.
All results are summarized in Table~\ref{tab:main-eval}.

\paragraph{Key finding.}
Across models, functional pass rates are substantially higher than security pass rates.
This pattern shows that when models are given only functional requirements, they often do not naturally implement protections that are later checked by the security harnesses.
Moreover, security scores have weak correlation with general code generation capability, which suggests that security compliance is not fully explained by overall coding strength.

\paragraph{Effect of iterative refinement.}
With collaborator feedback, models resolve functional issues quickly and typically reach a high level of functional correctness after about one to two refinement iterations on average.
However, this rapid functional improvement yields little gain in security performance.
Because the refinement loop is driven by functional failures, it rarely addresses vulnerabilities that do not manifest as functional errors, so functional saturation does not translate into security compliance.

\paragraph{Robustness under larger sampling budgets.}
Pass@5 mainly improves functional success rates, while security scores for most models increase only marginally.
Across independent attempts, models often converge to functionally correct solutions that repeatedly exhibit similar security weaknesses, which suggests that the dominant failure mode is systematic rather than accidental, and requires security-specific evaluation.

\subsection{Impact of Prompt Explicitness}
\begin{figure}[t]
  \centering
  \begin{subfigure}[t]{0.50\columnwidth}
    \centering
    \includegraphics[width=\linewidth]{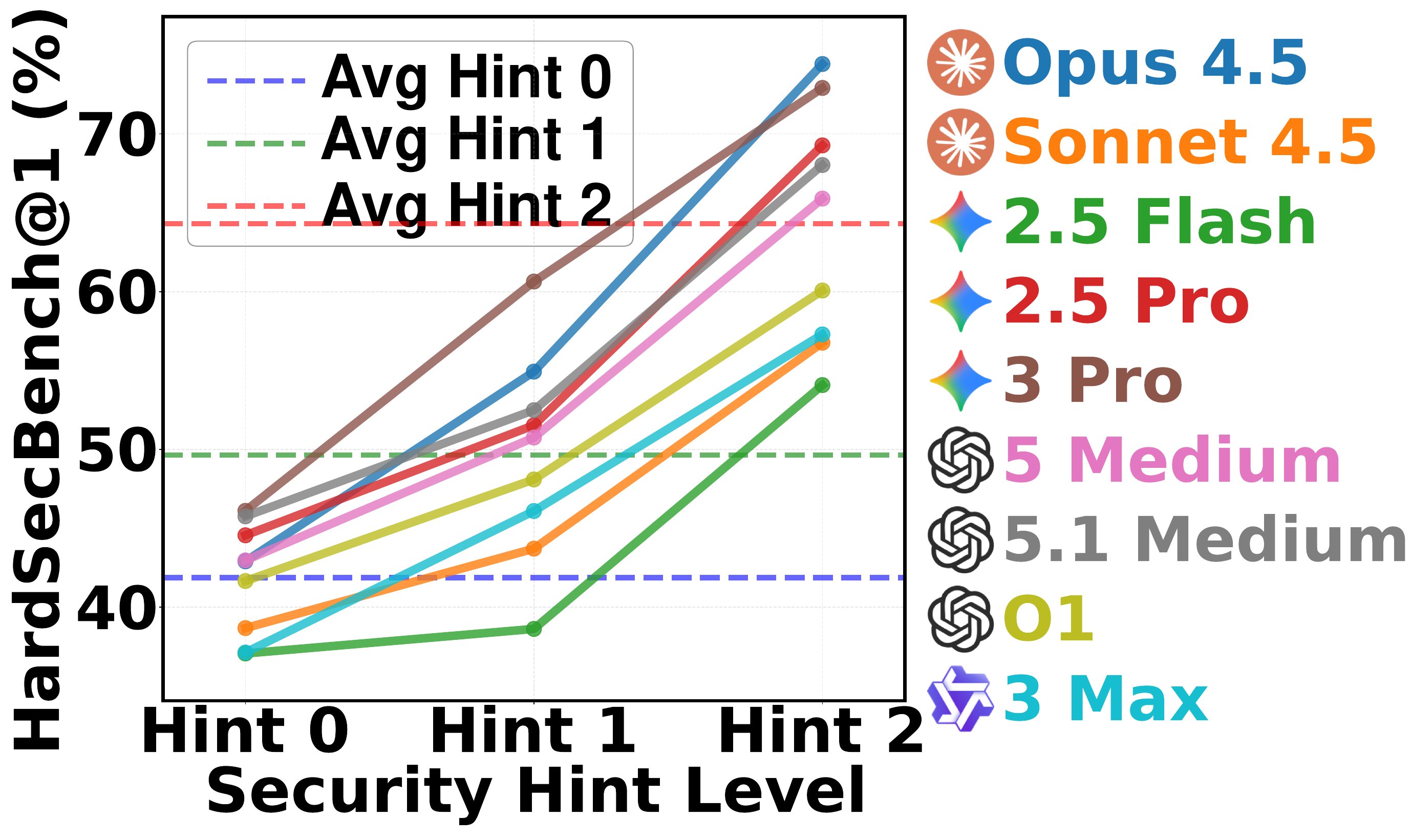}
    \caption{Closed-source models.}
    \label{fig:sub-a}
  \end{subfigure}\hfill
  \begin{subfigure}[t]{0.50\columnwidth}
    \centering
    \includegraphics[width=\linewidth]{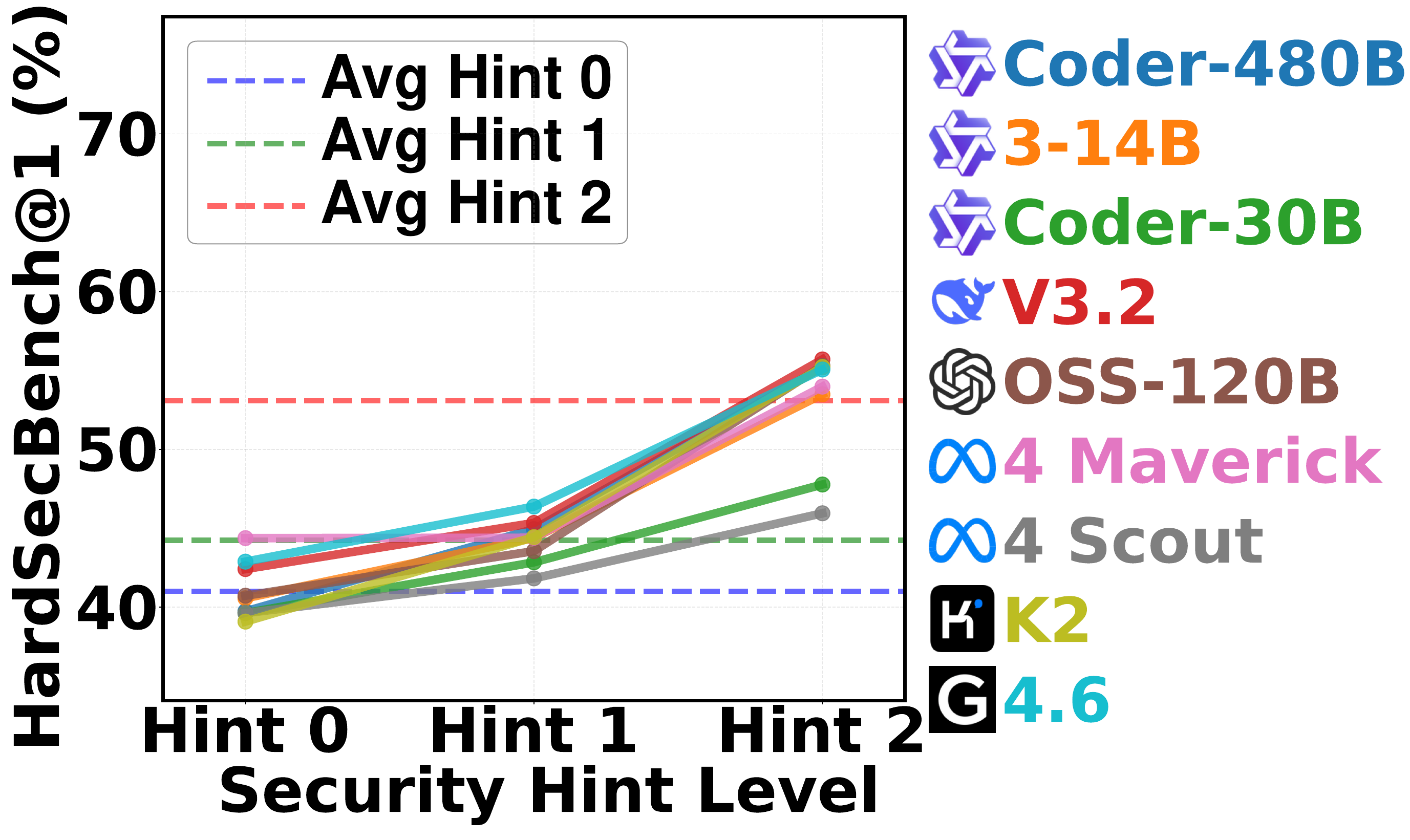}
    \caption{Open-source models.}
    \label{fig:sub-b}
  \end{subfigure}
\caption{Security pass rates across different security hint levels for representative models.~(a) Closed-source models,~(b) Open-source models. Model label colors indicate the corresponding lines.}

  \label{fig:hint}
\end{figure}

We examine whether security failures mainly arise from limited security knowledge or from weak security awareness. To this end, we conduct a prompt sensitivity analysis by varying the explicitness of security guidance across three levels. Hint 0 includes only functional requirements, Hint 1 adds general security reminders, and Hint 2 explicitly mentions specific vulnerability classes. Figure~\ref{fig:hint} shows how these hints affect security relative to general code generation capability.

\paragraph{Activation of Security Knowledge and Model Differences.}
The results reveal a clear disparity between closed-source and open-source models in activating security knowledge. For several closed-source models, Hint 1 yields noticeable gains, showing that even general security reminders can shift behavior toward safer outputs; Hint 2 delivers larger improvements, implying that security expertise is present in training but remains dormant when only functional requirements are provided. By contrast, smaller open-source models remain weakly responsive even under Hint 2, suggesting that their security performance is primarily constrained by limited capability rather than unactivated knowledge.

\begin{figure}[t]
  \centering
  \begin{subfigure}[t]{0.48\linewidth}
    \centering
    \includegraphics[width=\linewidth]{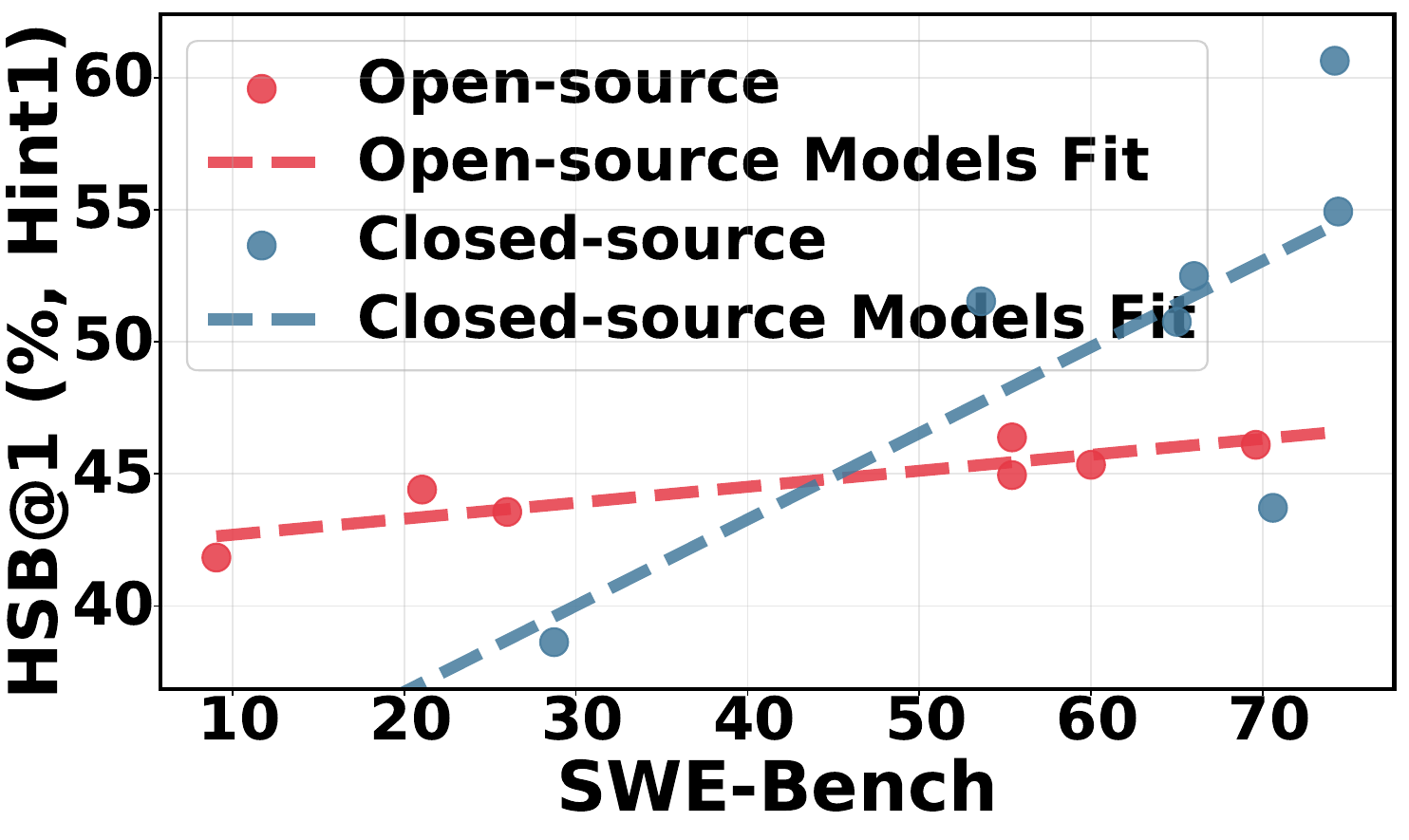}
    \caption{Hint 1 vs. SWE-bench.}
    \label{fig:swe-hint1}
  \end{subfigure}\hfill
  \begin{subfigure}[t]{0.48\linewidth}
    \centering
    \includegraphics[width=\linewidth]{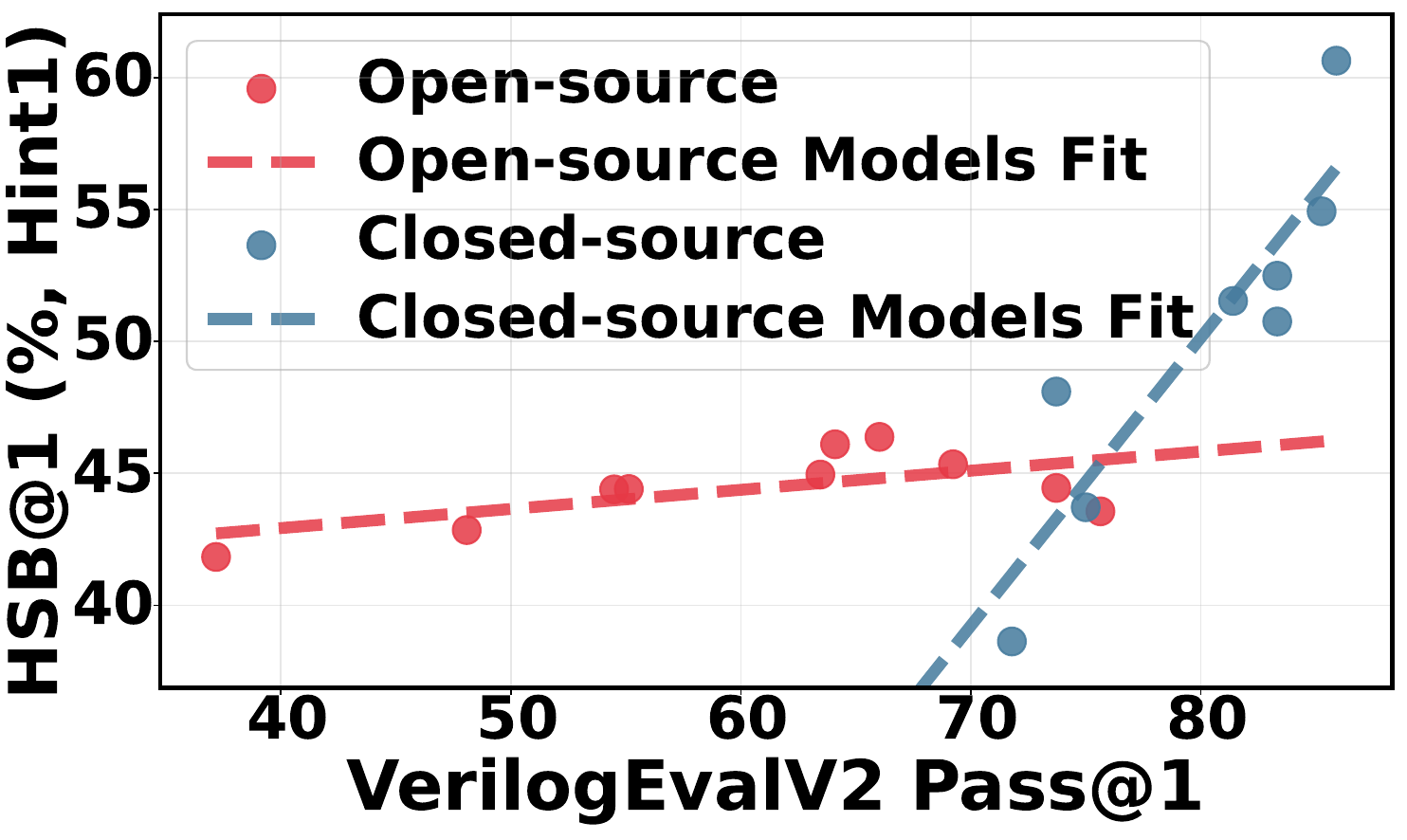}
    \caption{Hint 1 vs. VerilogEvalV2.}
    \label{fig:verilog-hint1}
  \end{subfigure}

  \begin{subfigure}[t]{0.48\linewidth}
    \centering
    \includegraphics[width=\linewidth]{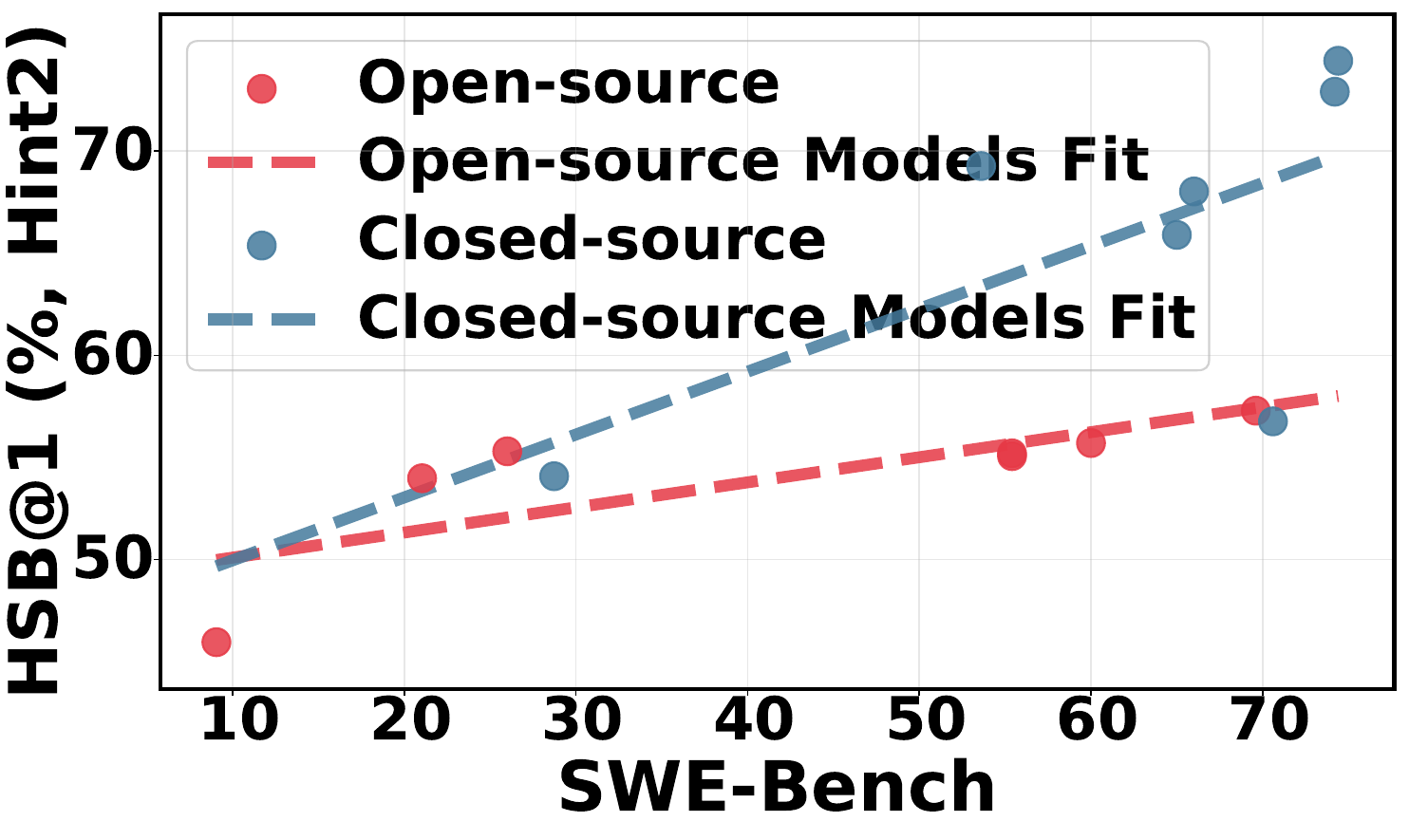}
    \caption{Hint 2 vs. SWE-bench.}
    \label{fig:swe-hint2}
  \end{subfigure}\hfill
  \begin{subfigure}[t]{0.48\linewidth}
    \centering
    \includegraphics[width=\linewidth]{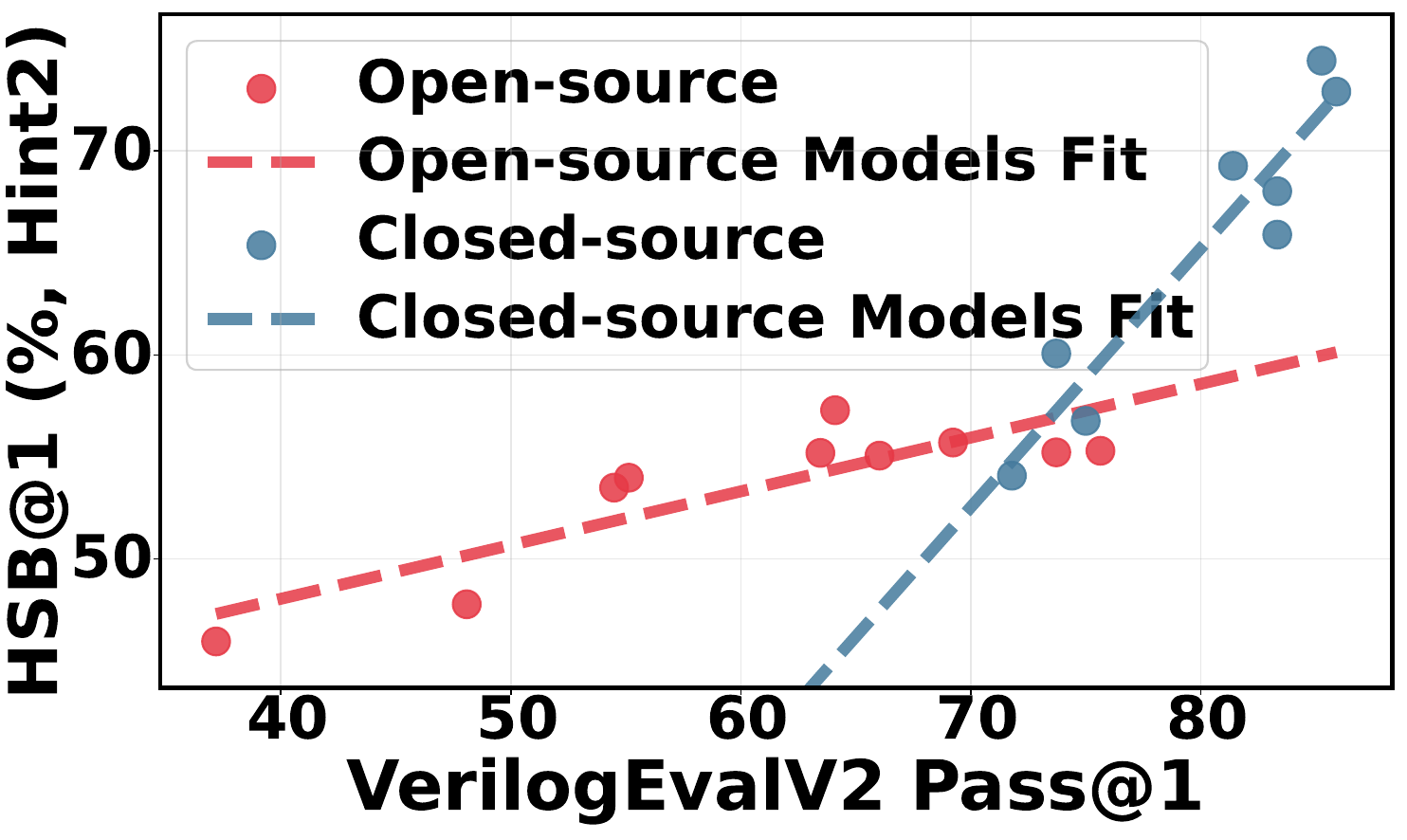}
    \caption{Hint 2 vs. VerilogEvalV2.}
    \label{fig:verilog-hint2}
  \end{subfigure}

 \caption{Correlation between code generation capability and security potential across prompting levels.}
  \label{fig:correlation-analysis}
\end{figure}

\paragraph{Relationship Between Security Potential and General Coding Capability.}
Figure~\ref{fig:correlation-analysis} shows a strong positive correlation between general code generation capability and security performance under explicit guidance. Under the Hint~0 setting, security pass rates show no significant correlation with general code generation capability. In contrast, under Hint~1 and Hint~2, security performance increases with general coding capability. As general coding capability increases, closed-source models exhibit a larger increase in security potential than open-source models.
This pattern suggests that security improvements under guidance are closely related to general code generation capability, especially for models with stronger reasoning capacity.

\subsection{Impact of Domain-Specific Fine-tuning}

We evaluate whether hardware-specialized fine-tuning reduces security gaps relative to the corresponding base models. Our analysis of CodeV-R1~\cite{zhu2025qimengcodevr1} and RTLCoder~\cite{10720939} shows that both series improve security performance on the Verilog subset of $\text{HardSecBench}_{\text{1-attempt}}@1$ after domain-specific training. By comparison, CodeV-R1 shows substantially larger gains, with clear improvements in both functional correctness and security compliance as prompts provide more explicit security guidance, whereas RTLCoder exhibits only marginal security improvements despite higher functional pass rates. This suggests that as security requirements become more complex, weaker models may struggle to follow the specification, resulting in lower functional correctness and security pass rates.

\begin{figure}[t]
  \centering
  \begin{subfigure}[t]{0.49\linewidth}
    \centering
    \includegraphics[width=\linewidth]{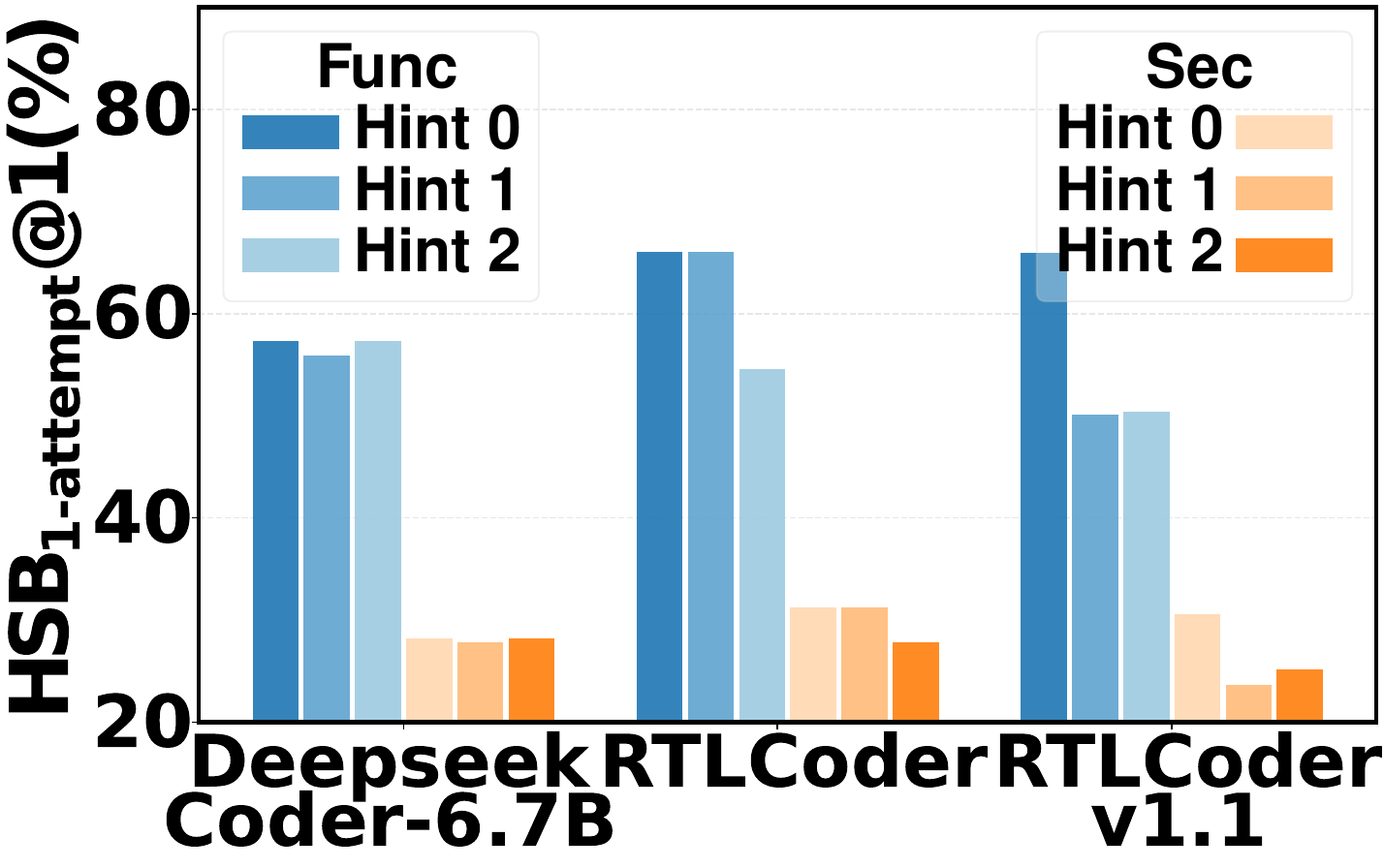}
    \caption{RTLCoder}
    \label{fig:qwen-specialized}
  \end{subfigure}\hfill
  \begin{subfigure}[t]{0.49\linewidth}
    \centering
    \includegraphics[width=\linewidth]{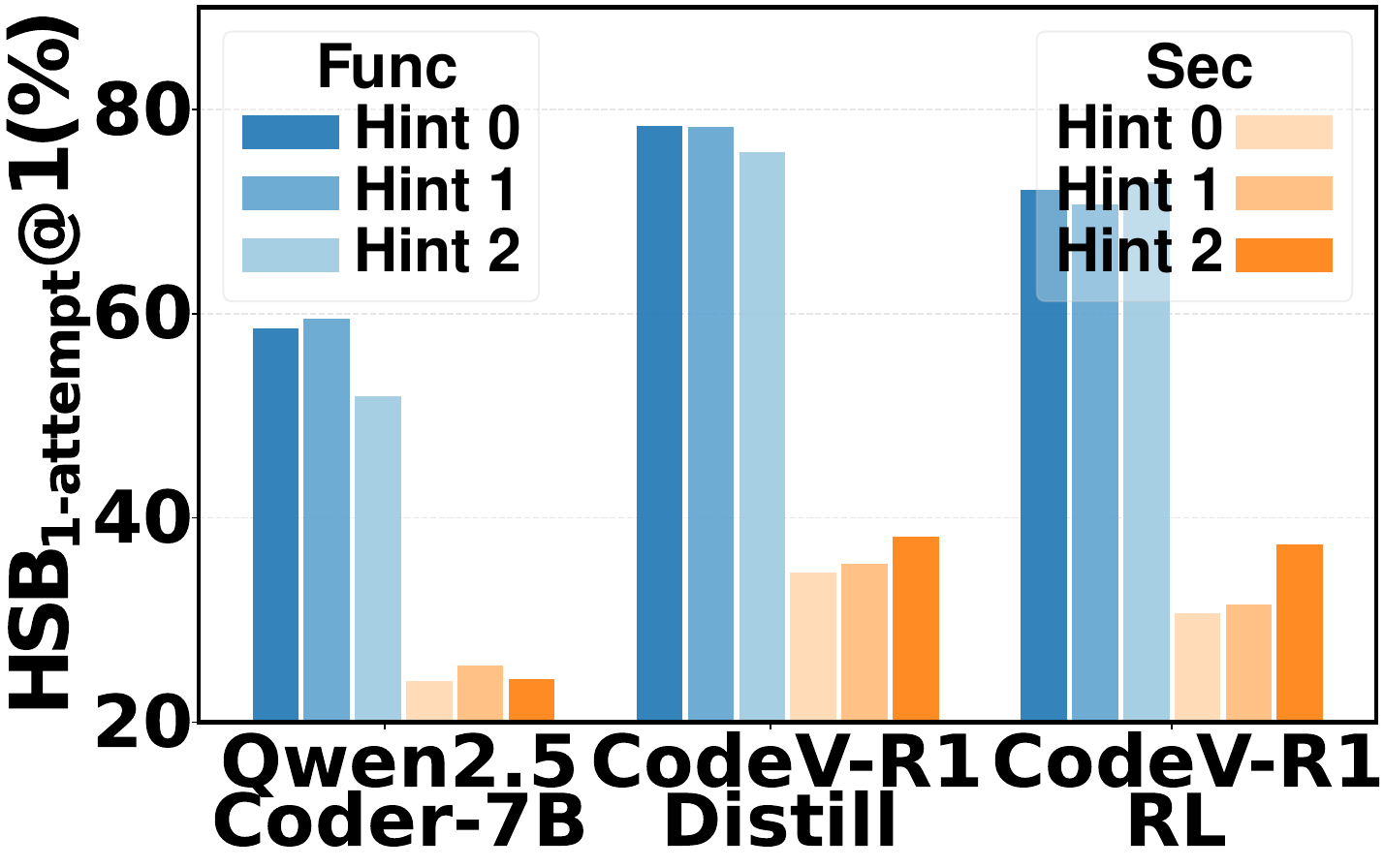}
    \caption{CodeV-R1}
    \label{fig:deepseek-specialized}
  \end{subfigure}
  \caption{Comparison of functional and security performance for hardware-specialized models across multiple prompting levels.}
  \label{fig:specialized-comparison}
\end{figure}

\paragraph{Specialized Fine-Tuning under Increasing Security Hints.}

Domain-specific fine-tuning improves the security baseline at Hint 0 compared to the base models. As security guidance becomes more explicit, the two model families diverge: CodeV-R1 exhibits consistent gains from Hint 0 through Hint 2 in both functional and security pass rates, while RTLCoder peaks at Hint 1 and drops at Hint 2, where highly explicit guidance reduces both functional and security performance. Further progress therefore requires stronger base models, higher-quality training data, and improved training recipes to better internalize secure behavior rather than merely follow explicit hints.

\subsection{Granular Analysis of CWE Domains}
We investigate the causes of security failures through a fine-grained evaluation over 76 hardware-related sub-CWEs. We follow the CWE hierarchy to map these sub-CWEs to 12 higher-level categories, and we consider two domains: firmware C and RTL Verilog implementations. Table~\ref{tab:cwe_mapping} summarizes the mapping and identifiers.

\begin{figure}[t]
  \centering
  \begin{subfigure}[t]{0.49\columnwidth}
    \centering
    \includegraphics[width=\linewidth]{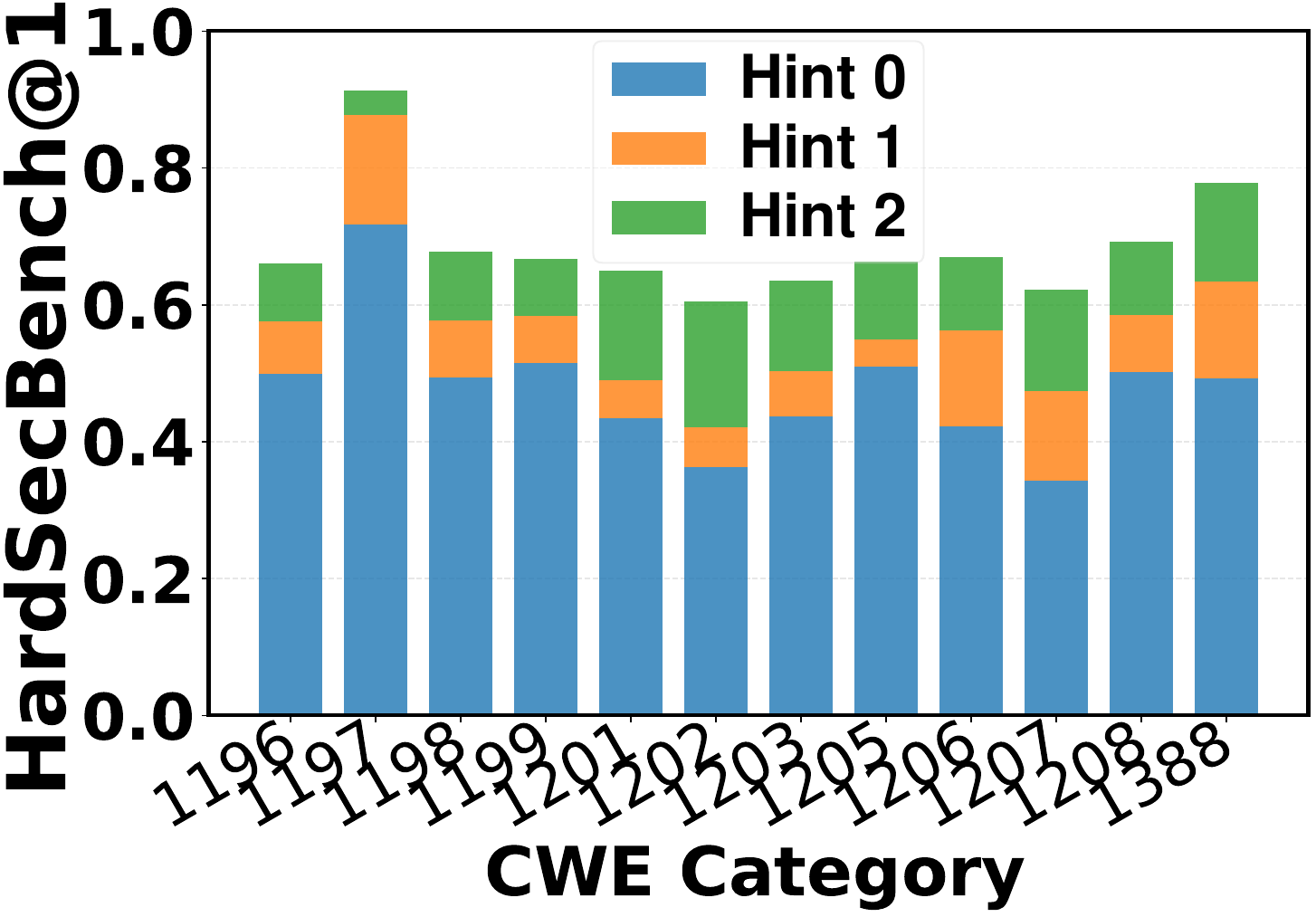}
    \caption{Closed-source models.}
    \label{fig:cwe-pass-rates-a}
  \end{subfigure}\hfill
  \begin{subfigure}[t]{0.49\columnwidth}
    \centering
    \includegraphics[width=\linewidth]{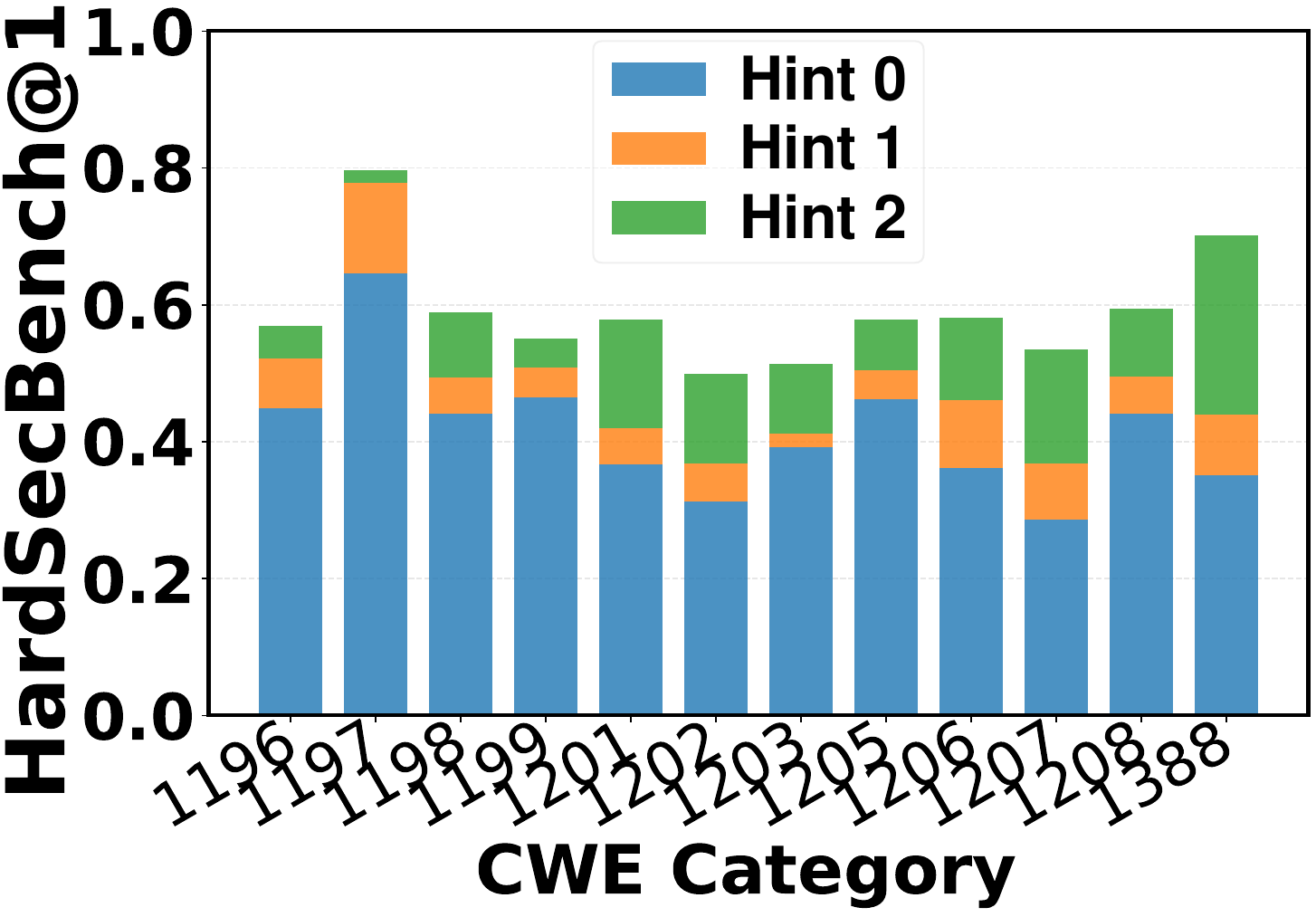}
    \caption{Open-source models.}
    \label{fig:cwe-pass-rates-b}
  \end{subfigure}
  \caption{CWE pass rates for different model categories:~(a) closed-source models,~(b) open-source models}
   \vspace{-5pt}

  \label{fig:cwe-pass-rates}
\end{figure}

\begin{table}[ht]
\centering
\small
\begin{tabular}{m{1cm} m{6.5cm}}
\toprule
\textbf{CWE} & \textbf{Category Name} \\
\midrule
1196 & Security Flow Issues \\
1197 & Integration Issues \\
1198 & Privilege Separation and Access Control Issues \\
1199 & General Circuit and Logic Design Concerns \\
1201 & Core and Compute Issues \\
1202 & Memory and Storage Issues \\
1203 & Peripherals, On-chip Fabric, and Interface/IO Problems \\
1205 & Security Primitives and Cryptography Issues \\
1206 & Power, Clock, Thermal, and Reset Concerns \\
1207 & Debug and Test Problems \\
1208 & Cross-Cutting Problems \\
1388 & Physical Access Issues and Concerns \\
\bottomrule
\end{tabular}
\caption{Aggregation of Evaluated Hardware sub-CWEs into Primary Categories.}
\vspace{-5pt}
\label{tab:cwe_mapping}
\end{table}

\paragraph{Firmware and RTL Domain Analysis.}
Comparing firmware and hardware tasks reveals only a small difference in baseline security awareness. On firmware vulnerabilities written in C, models achieve a security pass rate of 38.31\%. On hardware tasks in Verilog, the security pass rate is 44.93\%, a gap of 6.63 percentage points. Overall, both domains show similarly weak security awareness, because key security checks are not met reliably. The small domain gap suggests that security failures are not driven by language syntax, but by limited ability to recognize potential risks in system behavior. Improving performance will require training that strengthens this risk awareness across both C firmware and RTL designs.

\paragraph{Security Results by Vulnerability Category.}
Figure~\ref{fig:cwe-pass-rates} shows substantial variation across vulnerability categories, defined in Table~\ref{tab:cwe_mapping}. Models achieve higher baseline security pass rates on integration issues and on issues involving physical access. Performance is lowest on categories related to power, clock, thermal, and reset, as well as memory and storage. Many failures reflect difficulties with temporal behavior, including state machine consistency and interactions across clock domains, and they also reflect incomplete handling of physical access considerations. The impact of stronger hints depends on the model group as shown in Figure~\ref{fig:cwe-pass-rates}. Closed-source models improve notably on physical access under Hint 2. Open-source models show limited gains on categories involving peripherals and interface and I/O requirements, which suggests weaker coverage of these areas.

\section{Conclusion}
We propose HardSecBench, a benchmark for evaluating security awareness in hardware code generation.
HardSecBench uses a multi-agent pipeline to generate samples; for each task, it produces a structured specification, a secure reference implementation, and requirement-level harnesses. The implementation and harnesses are produced in separate contexts and validated with simulation evidence to reduce coupling.
Our experiments show that strong functional pass rates often coexist with missing security protections.
Prompting effects differ across model families, suggesting that some models require explicit guidance to surface security behavior while others lack it more fundamentally.
These findings highlight pressing challenges and offer actionable insights for future advancements in LLM-assisted hardware design.

\section*{Acknowledgments}
This work was supported by the Anhui Provincial Science and Technology Innovation Key Research Program under Grant No.202523J08050016 and the ZTE Industry-University-Institute Cooperation Funds under Grant No.IA20250000009.

\bibliographystyle{named}
\bibliography{ijcai26}

\end{document}